\documentstyle[12pt]{article}

\advance \textheight by \topskip
\begin{document}

\title{\bf Comment on
``The relativistic particle with curvature and torsion
of world trajectory"}

\author{
Mikhail S. Plyushchay\thanks{E--mail:
  mplyushc@lauca.usach.cl}\\
{\small \it Departamento de F\'{\i}sica,
Universidad de Santiago de Chile}\\
{\small \it Casilla 307, Santiago 2, Chile}\\
{\small and}\\
{\small \it Institute for High Energy Physics,
  Protvino, Moscow region, Russia}}
\date{}

\maketitle
\vskip-1.0cm

\begin{abstract}
Gogilidze and Surovtsev have claimed recently (hep-th/9809191)
that the tachyonic sector can be removed from the spectrum of
the relativistic particle with curvature and torsion by a proper
gauge choice.  We show that the mass-spin dependence obtained by
them is incorrect and point out that their gauge
surface does not cross all the gauge orbits.  We discuss the
nature of the tachyonic sector of the model and argue why it
cannot be removed by any gauge fixing procedure.
\end{abstract}

The model of the relativistic particle with curvature and torsion is
given by the action
\begin{equation}
A=-\int(m+\alpha k+\beta\kappa)ds,\quad
k^2=x''{}^2,\quad \kappa=\epsilon^{\mu\nu\lambda}
x'_\mu x''_\nu x'''_\lambda/k^2,
\label{1}
\end{equation}
where $ds^2=-dx^\mu dx_\mu$, $x'_\mu=dx_\mu/ds$, $k$ and
$\kappa$ are the curvature and torsion of the particle's world
trajectory, $m$ is a mass parameter, $\alpha$ and $\beta$ are
dimensionless parameters.
In a
given parametrization  $x_\mu=x_\mu(\tau)$ action (1)
takes the form
\begin{equation}
A=\int Ld\tau,
\quad L=-\sqrt{-\dot{x}{}^2}(m +\alpha k+\beta\kappa)
\label{2}
\end{equation}
with
$k=\sqrt{(\ddot{x}{}^2\dot{x}^2-(\ddot{x}\dot{x})^2)/
(\dot{x}{}^2)^{3}}$ and
$\kappa=-
\epsilon^{\mu\nu\lambda}\dot{x}_\mu
\ddot{x}_\nu \stackrel{...}
{x}_\lambda/(\ddot{x}{}^2\dot{x}{}^2
-(\ddot{x}\dot{x})^2)$.
This model was investigated in detail in ref.\cite{kp1} both at
the classical and quantum levels.  It was
shown that the model has classical and quantum solutions
characterized by the timelike, lightlike and spacelike
energy-momentum vector.  The classical
spin-mass dependence in the massive and tachyonic
sectors is given by (see eqs. (2.13) and (2.14) in \cite{kp1})
\begin{equation}
\sqrt{|p^2|}=m\frac{\beta S+\alpha\sqrt{S^2+\theta(\alpha^2
-\beta^2)}}{S^2+\theta\alpha^2}>0,
\label{sm}
\end{equation}
where $p_\mu$ and $S$ are the energy-momentum vector
and the spin of the system and $\theta=+1$ and $\theta=-1$
correspond to the massive ($-p^2>0$) and tachyonic ($-p^2<0$)
states. According to (\ref{sm}),
in the sector $\theta=+1$
the mass of the states turns out to be
restricted from above by the
value of the parameter $m$ and $-p^2\propto S^{-2}$ for $S^2>>1$.
These properties of the massive states
together with the fact of the presence of the massless and tachyonic
sectors are the characteristic properties of the spectrum of the
infinite-component Majorana equation (see \cite{kp1} and references
therein).

Recently this model has been studied at the classical level
by Gogilidze and Surovtsev \cite{gs}.
They have investigated local symmetries of action (\ref{2}),
have obtained another, ``two-branched expression" \cite{nest}
for the  dependence between mass and
spin\footnote{In \cite{gs}
the metric and the parameter $\beta$
are opposite in sign to those chosen here.}
\begin{equation}
S=\pm\alpha\sqrt{(m^2/|p^2|)-\theta}
+\beta m/\sqrt{|p^2|},
\label{ngs}
\end{equation}
and claimed that the tachyonic sector can be removed by a proper
gauge choice.  In this Comment we show that though
the ``two-branched" dependence (\ref{ngs})
works well for some choice of the parameters $\alpha$ and $\beta$,
it is not correct in general case, and argue why the claim on the
possibility to remove tachyonic sector
by the choice of gauge is untrue.

First, let us note that (\ref{2}) is obtained from the
reparametrization-invariant form of action
(\ref{1}).  As a consequence, the
invariance of action (\ref{2})
under reparametrization transformations
$\delta x_\mu=\lambda(\tau)\dot{x}_\mu$ is obvious
and does not require special methods used in \cite{gs}
to derive this symmetry.

Lagrangian equations of motion of the
higher-derivative system (\ref{2})
have the form $\frac{d}{d\tau}p_\mu=0$,
where $p_\mu$ is the Noether
energy-momentum vector \cite{kp1}.
The vectors
$e_\mu=\dot{x}_\mu/\sqrt{-\dot{x}{}^2}=x'_\mu$,
$n_\mu=\dot{e}_\mu/\sqrt{\dot{e}{}^2}=x''_\mu/\sqrt{x''{}^2}$
and
$r_\mu=\epsilon_{\mu\nu\lambda}e^\nu n^\lambda$
form the orthonormal oriented basis
in (2+1)-dimensional space-time.
The contraction of the equations of motion
with $n_\mu$ gives rise to the lagrangian constraint
\cite{kp1}
\begin{equation}
(m-\beta\kappa)k-\alpha\kappa^2=0.
\label{curtor}
\end{equation}
This relation together with relation $\dot{k}=0$
following from eq. (\ref{curtor}) as a secondary lagrangian constraint
mean that the curvature and torsion of the worldline
are integrals of motion.
There are two special cases.
When $\alpha=0$, that corresponds to the model
of the relativistic particle with torsion \cite{tor},
the worldline's torsion is fixed,
$\kappa=m/\beta$, whereas curvature can take
arbitrary values on the halfline, $k\geq 0$.
Another special case is given by  $\beta=0$
and corresponds to the model of the relativistic
particle with curvature (or `rigidity') \cite{rig}.
Since the curvature is non-negative by definition,
relation (\ref{curtor}) means
that in this case the model has
nontrivial solutions only for $\alpha>0$.
As it was shown in \cite{rig},
in the case $\alpha<0$, $\beta=0$ the model admits only a
trivial motion
with $k=0$ corresponding to the motion of the
massive relativistic particle
($\alpha=\beta=0$).
Using the complete set of vectors,
the expression for the energy-momentum vector
can be reduced to \cite{kp1}
\begin{equation}
p_\mu=me_\mu+(\alpha\kappa+\beta k)r_\mu.
\label{pmu}
\end{equation}
For
$\alpha=0$ from (\ref{pmu})
we find the mass-curvature dependence:
$-p^2=m^2-\beta^2 k^2$. Therefore, in this case the
model has timelike  (for $0\leq k^2<m^2/\beta^2$),
lightlike ($k^2=m^2/\beta^2$) and
spacelike ($k^2>m^2/\beta^2$) solutions.
If $\alpha\neq0$, relation (\ref{curtor})
gives the mass-torsion dependence
\begin{equation}
-p^2=m^2(1-\alpha^2\kappa^2(m-\beta\kappa)^{-2}),\quad
{\rm sign}\, \alpha(m-\beta\kappa)>0,
\label{mtor}
\end{equation}
where the inequality follows from eq. (\ref{curtor}) and imposes
restrictions on the range of possible values of the torsion.
{}From these relations one finds that in general case, again,
there are massive ($-p^2>0$), massless ($p^2=0$) and tachyonic
($-p^2<0$) solutions in the system.  {\it Since the
energy-momentum vector is a gauge-invariant quantity, it is
clear that the spectrum of the model (the value of $p^2$)
cannot depend in any way on the gauge choice and, in particular,
the tachyonic sector cannot be removed from the spectrum of the
system by the gauge fixing procedure.  This means that if some
set of gauge-fixing conditions removes tachyonic sector (as ref.
\cite{gs} claims), the corresponding gauge surface does not
cross all the gauge orbits and, so, the choice of such
conditions as a gauge is not admissible}.  Moreover, let us note
that according to eq. (\ref{mtor}) the model has only
tachyonic solutions for $\alpha<0$, $\alpha^2\geq\beta^2$.

For $\alpha\neq0$ relation (\ref{curtor}) can be considered
as quadratic equation defining the worldline's torsion as
a function of its curvature:
$\kappa=\kappa_\pm(k)
=\frac{1}{2\alpha}(-\beta k\pm\sqrt{\beta^2k^2+4\alpha mk})$.
Substituting this in relation (\ref{mtor}),
we get the mass-curvature relation
\begin{equation}
-p^2=m^2-\frac{1}{4}\left(-\beta k\pm\sqrt{\beta^2k^2+4\alpha mk}
\right)^2
\label{mcur}
\end{equation}
corresponding to the relation $-p^2=m^2(1-k^2\kappa^{-2})$ which,
in turn, follows
from eqs. (\ref{curtor}) and (\ref{mtor}).
Generally the two different values of torsion, $\kappa_+(k)$ and
$\kappa_-(k)$, correspond to one value of curvature. Therefore,
{\it the general claim of ref. \cite{gs} that at a given value
of $k$ the particle has a mass corresponding either to upper or
lower sign in (\ref{mcur}) is incorrect}.

The Noether integrals corresponding
to the Lorentz symmetry are given by the vector
dual to the
angular momentum tensor of the system  \cite{kp1}:
$
{\cal J}_\mu=-\epsilon_{\mu\nu\lambda}x^\nu p^\lambda
+J_\mu,$
$J_\mu=\alpha r_\mu-\beta e_\mu.
$
For $p^2\neq0$, the particle spin  is defined
as $S={\cal J}p/\sqrt{|p^2|}=pJ/\sqrt{|p^2|}$.
It can be presented as a single-valued function of
the worldline's torsion \cite{kp1}.
The
corresponding spin-torsion relation together with eqs.
(\ref{curtor}), (\ref{pmu})
lead finally to the single-valued mass-spin relation
given  by eq. (\ref{sm}).
This relation itself specifies correctly the
region of possible values of spin \cite{kp1}
and `feels' the singular nature of the parameter region $\beta=0$,
$\alpha<0$.
Proceeding from  relation (\ref{sm})
one can reproduce the properties of the spectrum
discussed above. In particular,
according to this relation the system
has no timelike solutions in the case
$\alpha<0$, $\alpha^2\geq\beta^2$.

One could invert the relation
(\ref{sm}) and present spin as a function of $|p^2|$.
The result will be given by eq. (\ref{ngs}).
However, realizing such trivial
manipulations, it is necessary to
take a square of some algebraic relation
in order to obtain
the quadratic equation for spin as a function of mass.
As a result,
inverse spin-mass relation (\ref{ngs})
may contain some superfluous points not satisfying
the initial relation (\ref{sm}).
This indeed happens here:
the relation (\ref{ngs}) itself
does not exclude
timelike solutions ($\theta=+1$)
for $\alpha<0$, $\alpha^2\geq\beta^2$.
Moreover, it does not `feel'
the peculiar nature of the case $\beta=0$,
$\alpha<0$.
Thus, generally {\it the ``two-branched expression" (\ref{ngs})
of ref. \cite{gs} is incorrect}.

In conclusion, let us give some additional comments
on the tachyonic sector of the model.
In ref. \cite{kp1} the general solutions to the classical
equations of motion were constructed explicitly
in all the three mass sectors of the system.
It was also shown that the corresponding
solutions in the massive and tachyonic sectors
can be related by
the change $\sqrt{|p^2|}\rightarrow
i\sqrt{|p^2|}$ and that the massless solutions
may be obtained from the massive and tachyonic solutions
via the limit transition $p^2\rightarrow0$.
In this sense the massive, massless and tachyonic sectors
of the model (\ref{1}) are analogous to the
elliptic, parabolic and hyperbolic trajectories
of the Kepler problem.

The relation of the model (\ref{1}) to the quantum Majorana
equation having massive, massless and tachyonic
solutions was observed for the first time
for the case of the relativistic particle with curvature ($\beta=0$)
\cite{rig}. Lately, in ref. \cite{tor}
it was shown that the equation
describing the quantum states of the
model of the relativistic particle
with torsion ($\alpha=0$) is exactly the
(2+1)-dimensional analog of the
Majorana infinite-component linear differential
equation.
This means that the model of the relativistic particle
with torsion is the classical analog
for the quantum system described by the Majorana equation.
In ref. \cite{dual} it was observed the interesting duality
between the (2+1)-dimensional
relativistic models of the particle with torsion
and the massive scalar particle with charge $q$ in a background
of the homogeneous constant electromagnetic field.
The duality relation has the form
$\beta^{-1}p^\mu\leftrightarrow qm^{-1}R^\mu$
and establishes the identity of the classical
evolution of the particles in both systems.
Here $p^\mu$ is the dynamical energy-momentum
vector for the first system and $R^\mu=(H,E^2,-E^1)$
is the vector dual to the electromagnetic
field tensor in the second system.
Therefore, from the point of view of this duality
the massive, massless and tachyonic solutions
of the relativistic particle with torsion
correspond to the three possible `configurations'
($-R^2>0$, $R^2=0$ and $-R^2<0$)
of the electromagnetic field to which
the charged scalar particle is minimally coupled.

Therefore, {\it the tachyonic sector
in model (\ref{1}) is not the gauge artifact,
but is its characteristic property.}

At last, let us note that the presence of
the tachyonic sector is not the necessary feature of the
geometric models of relativistic particles with higher
derivatives.  The first exceptional case is the
(3+1)-dimensional model of the massless particle with rigidity
\cite{m0}, which classically reveals $W$-symmetry \cite{ram}
and at the quantum level
describes massless bosons and fermions \cite{m0,ram}.
Other tachyonless models were found in the
class of (2+1)-dimensional systems characterized by
the action $A=-\int(m+kf(\xi))ds$, $\xi=k/\kappa$, to
which the model (\ref{1}) belongs (see ref. \cite{kp2}).

\vskip0.3cm
The work was supported in part
by grant 1980619 from FONDECYT (Chile).

\end{document}